\documentclass[aps,pre,twocolumn,floats]{revtex4}
\usepackage{array}
\usepackage{epsfig}
\usepackage{bm}
\usepackage{graphicx}% Include figure files
\usepackage{dcolumn}% Align table columns on decimal point

\begin{document}

% basic commands
\newcommand{\hide}[1]{}
\newcommand{\tbox}[1]{\mbox{\tiny #1}}
\newcommand{\half}{\mbox{\small $\frac{1}{2}$}}
\newcommand{\sinc}{\mbox{sinc}}
\newcommand{\const}{\mbox{const}}
\newcommand{\trc}{\mbox{trace}}
\newcommand{\intt}{\int\!\!\!\!\int }
\newcommand{\ointt}{\int\!\!\!\!\int\!\!\!\!\!\circ\ }
\newcommand{\eexp}{\mbox{e}^}
\newcommand{\bra}{\left\langle}
\newcommand{\ket}{\right\rangle}
\newcommand{\EPS} {\mbox{\LARGE $\epsilon$}}
\newcommand{\ar}{\mathsf r}
\newcommand{\im}{\mbox{Im}}
\newcommand{\re}{\mbox{Re}}
\newcommand{\bmsf}[1]{\bm{\mathsf{#1}}}

%%%%%%%%%%%%%%%%%%%%%%%%%%%%%%%%%%%%%%%%%%%%%%%%%%%%%%%%%%%%%%%%

\title{Analysis of airplane boarding via space-time geometry and random matrix theory}

\author{Eitan Bachmat$^{1}$, Daniel berend$^{1,2}$, Luba Sapir$^{3}$, Steven Skiena$^{4}$, Natan Stolyarov$^{5}$ }

\affiliation{
$^{1}$ Department of Computer Science, Ben-Gurion University, Beer-Sheva 84105, Israel \\
$^{2}$ Department of Mathematics, Ben-Gurion University, Beer-Sheva 84105, Israel \\
$^{3}$ Department of Management and Industrial Engineering, Ben-Gurion University, Beer-Sheva 84105, Israel \\
$^{4}$ Department of Computer science, SUNY at Stony Brook, Stony Brook NY, 11794, USA \\
$^{5}$ Department of computer science, Ben-Gurion University, Beer-Sheva 84105, Israel
}

%%%%%%%%%%%%%%%%%%%%%%%%%%%%%%%%%%%%%%%%%%%%%%%%%%%%%%%%%%%%%%%%

\begin{abstract}
We show that airplane boarding can be asymptotically modeled by
2-dimensional Lorentzian geometry. Boarding time is given by the maximal
proper time among curves in the model. Discrepancies between the model
and simulation results are closely related to random matrix theory. We
then show how such models can be used to explain why some commonly practiced
airline boarding policies are ineffective and even detrimental.
\end{abstract}

\maketitle

Airplane boarding is a process experienced daily by millions of passengers
worldwide. Airlines have developed various strategies in the hope of
shortening boarding time, typically leading to announcements of the form
``passengers from rows 40 and above are now welcome to board the
plane'', often heard around airport terminals. We will show how the
airplane boarding process can be asymptotically modeled by spacetime
geometry. The discrepancies between the asymptotic analysis and finite
population results will be shown to be closely related to random matrix
theory (RMT). Previously, airplane boarding has only been analyzed via
discrete event simulations \cite{MMM,VB,VVH}. 

We model the boarding process as follows: Passengers $1,...,N$ are
represented by coordinates $X_i=(q_i,r_i)$, where $q_i$ is the index of
the passenger along the boarding queue (1st, 2nd, 3rd and so on),
and $r$ is his/her assigned row number. 
We rescale $(q,r)$ to $[0,1]\times [0,1]$.
It is assumed that the main cause of delay in airplane boarding is the
time it takes passengers to organize their luggage and seat themselves
once they have arrived at their assigned row. The input parameters for
our model are:

$u$ -- Average amount of aisle length occupied by a passenger.

$w$ -- Distance between successive rows. 
%For the purposes of presentation we will assume that $w$ is fixed.

$b$ -- Number of passengers per row.

$D$ -- Amount of time (delay) it takes a passenger to clear the aisle,
once he has arrived at his designated row. 
%We shall assume at first that
%$D$ is fixed.

$p(q,r)$ -- The joint distribution of a passenger's row and queue joining time.
$p(q,r)$ is directly affected by the airline policy and the way
passengers react to the policy.

For the purposes of presentation, we shall assume that $u,w,b,D$ are all
fixed. The airplane boarding process produces a natural partial order
relation of blocking among passengers. Passenger $X$ {\it blocks} passenger
$Y$ if it is impossible for passenger $Y$ to reach his assigned row
before passenger $X$ (and others blocked by $X$) has sat down and
cleared the aisle. Airplane boarding functions as a peeling process
for the partial order defined by the blocking relation. At first,
passengers who are not blocked by others sit down;
these passengers are the minimal elements under the blocking relation. In
the second round, passengers who are not blocked by passengers other
than those of the first round are seated, and so forth. Boarding time
thus coincides with the size of the longest chain in the partial order.

We assign to the boarding process with parameters $u,b,w,D,p(q,r)$ a
Lorentz metric defined on the $(q,r)$ unit square by 
\begin{equation}
\label{model}
ds^2=4D^2p(q,r)(dqdr+k\alpha (q,r)dq^2),
\end{equation}
where $k=bu/w$ and $\alpha (q,r)=\int_r^1p(q,z)dz$. 
There are two properties of the metric which relate it to the boarding process:
\begin{itemize}
\item{} The volume form of the metric is proportional to the passenger
density distribution $p(q,r)$.

\item{} The blocking partial order among passengers during the boarding
process asymptotically coincides with the past-future causal relation induced
by the metric on the passengers, viewed as events in space-time via their
$q,r$ coordinate representation. 
\end{itemize}

To establish the second property, consider passengers represented by
$X=(q,r)$ and $X'=(q+dq,r+dr)$, $dq>0$. Consider the time
passenger $X$ arrives at his designated row. All passengers with row
numbers beyond $r$, who are behind $X$ in the queue but in
front of $X'$, will occupy aisle space behind $X$.
The number of such passengers is roughly $N\alpha dq$. Each such
passenger occupies $u/w$ units of aisle length, where we take the basic
aisle length unit to be the distance between rows. The row difference
between $X$ and $X'$ is $-(N/b)dr$. We conclude that passenger $X$ is
blocking passenger $X'$, via the passengers which are behind him,
roughly when $dq\geq -\alpha kdr$, a condition which coincides
(together with $dq>0$) with the causal relation induced by the metric.
By the two main properties we may approximate asymptotically the
airplane boarding process by the peeling process applied to the
past-future causal relation on points in the associated spacetime,
sampled with respect to the volume form. By a well-known result,
2-dimensional Lorentzian metrics are conformally flat, and hence, after an
appropriate coordinate transformation, we may assume that the spacetime
is given by a metric of the form
\begin{equation}
ds^2=r(x,y)dxdy
\end{equation}
on some domain (not necessarily the unit square). In the
new coordinates $x,y$, which are lightlike coordinates, chains in the
causal relation coincide with increasing (upright) subsequences,
namely, sequences of points $(x_i,y_i)$ such that $x_i\leq x_j$ and
$y_i\leq y_j$ for $i<j$. The peeling process applied to the causal relation
coincides in this case with patience sorting which is a well-known
card game process, which 
computes the longest increasing subsequence in a permutation \cite{Ma,AD}.

%The global scaling constant $4D^2$ comes from a result 
%of Vershik and Kerov \cite{VK}. Statement (B) follows from statement (A) and the fact that the level 
%curves given by the equation $T(X)=\tau $ are Lorentz-orthogonal to maximal proper time trajectories.
%A previous computation of Aldous and Diaconis on pile sizes in patience sorting \cite{AD} corresponds to statement (B)
%when the space-time is assumed to be flat.

Denote by $T(X)$ the maximal proper time (integral over $ds$) of a
timelike trajectory ending at $X=(q,r)$ and by $L(\tau)$ the length 
(integral over $\sqrt{-ds^2}$) of the spacelike curve defined 
by the equation $T(X)=\tau $. \newline
Using the analysis of the size of maximal increasing subsequences given in \cite{VK,DZ}, the two basic properties lead 
to the following modeling statements.

(A) The boarding time of passenger $X_i$ is approximately $\sqrt{N}T(X_i)$.
In particular, the total boarding time is approximately
$\sqrt{N}d$, where $d=Max_XT(X)$ is the maximal proper time of a
curve in the unit square with respect to the Lorentzian metric.

(B) Let $N(\tau)$ be the number of passengers with boarding time at
most $\sqrt{N}\tau$ and $\tilde{N} (\tau)=N(\tau)/N$. Then,
$d\tilde{N}(\tau )=\frac{1}{2D}L(\tau)d\tau$.

Here the word ''approximately" is used to mean that the ratio between the
two quantities tends to~$1$ with probability 1 as the number of passengers $N$
tends to infinity.

We apply statement (A) to the analysis of boarding times. Consider
first the case where the airline does not have a boarding policy,
namely, passengers queue at uniformly random times, so that
$p(q,r)=1$, $\alpha =1-r$ and the corresponding metric is
\begin{equation}
\label{nopolicy}
ds^2=4dq(dr+k(1-r)dq).
\end{equation}
We use this model to study the effect of airplane design parameters
such as the distance between rows and the number of passengers per row
on boarding time. These parameters affect boarding time through the
parameter $k=\frac{ub}{w}$. To find the maximal proper time curve, we solve the
Euler-Lagrange equation for proper time subject to the constraints of
lying in the unit square. In figure 1 the maximal proper time curve is
plotted for several values of the parameter $k$. For $k\leq \ln 2$ the
curve is contained in the interior of the square and is therefore a
geodesic. The length of the curve is
\begin{equation} \label{dk1}
d(k)=2\sqrt{\frac{e^k-1}{k}}\,.
\end{equation}
For $k>\ln 2$ the maximal
curve ``crawls'' at first along the $q$-axis until reaching a point
$B=(q(k),0)$ such that the geodesic between $B$ and $A=(1,1)$ has a
vanishing derivative at $B$. We have
\begin{equation}
d(k)=2\sqrt{k}+2(1-\ln 2)/\sqrt{k}\,.
\end{equation}
%Figure 2 shows the effect of $k$ on total boarding time.
%\begin{figure}[htbp]
%\centerline{\includegraphics[width=4.75in,height=2.25in]{kplot.eps}} \label{fig1}
%\end{figure}
%
%\begin{figure}
%\includegraphics[width=2.25in,height=2.25in]{kplot}% Here is how to import EPS art
%\caption{\label{fig:epsart} The dependence of total boarding time on $k$ when $p$ is uniform}
%\end{figure}
%
%Figure 3 compares the trail of pointers starting at the point $(1,1)$ with the limiting curve $C_A$ which was
%computed analytically. The pointers were obtained from simulations of the boarding process with parameter
%settings $k=0.5$.
%
%\begin{figure}
%\includegraphics[width=2.25in,height=2.25in]{eitan}% Here is how to import EPS art
%\caption{\label{fig:epsart} Maximal proper time curves for $p$ uniform and $k=0,\frac{1}{2},1,5$}
%\end{figure}
%HERE WAS PLACE FOR FIG 1 and 2
%\begin{figure}[htbp]
%\centerline{\includegraphics[width=4.75in,height=2.25in]{sheme.eps}} \label{fig1}
%\end{figure}
%
%We turn to the study of airline boarding policies. 
We compared the above computations with simulations of the boarding
process. In Figures 1 and 2 we show, for $k=0.5$ and $k=5$, the computed
maximal curve along with maximal chains obtained from simulations with
$N=100$, $N=200$ and $N=10^6$. As can be seen from the figures, when
$N=10^6$, the longest chain clusters along the computed curves. The
length of the maximal chain also matches well the estimated values
derived from the spacetime model. For $k=0.5$ the length of the
maximal chain in the simulation was $2261$, while the estimated value
is $2278$, while for $k=5$ the corresponding values were $4589$ and
$4740$, respectively. For the more realistic values of $N=100,200$
passengers, there are substantial deviations of the maximal chain from
the expected curve.
\begin{figure}
\includegraphics[width=2.25in,height=2.25in]{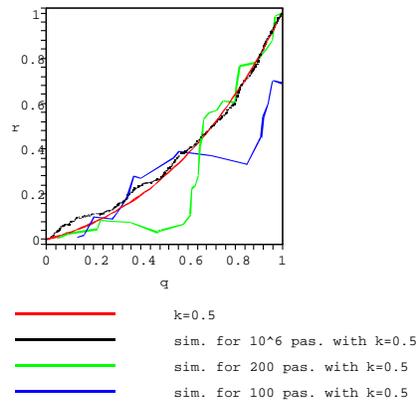}
% Here is how to import EPS art
\caption{\label{fig:epsart1} Comparison of the maximal curve for $k=0.5$
with simulation based maximal chains.}
\end{figure}
\begin{figure}
\includegraphics[width=2.25in,height=2.25in]{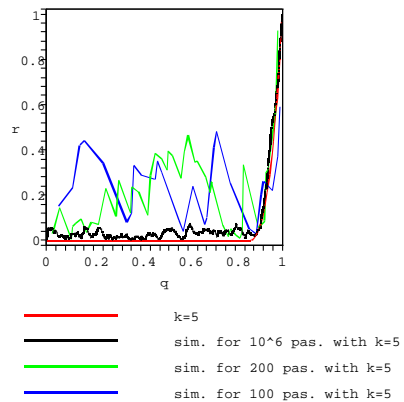}
% Here is how to import EPS art
\caption{\label{fig:epsart1} Comparison of the maximal curve for $k=5.0$
with simulation based maximal chains.}
\end{figure}
Table II presents the average boarding time results for 1000
simulations of the boarding process for several settings of $k$ and
$N$. We also provide the boarding time estimate computed via the
spacetime model. Upon inspection, Table II shows that, for realistic
values of $N$, there are substantial differences in the range of 20-60
percent between the asymptotic boarding time estimates computed via
Lorentzian geometry and the boarding time computed via simulations of
the boarding process. These large
differences correspond to the large deviations of the maximal chains
from the expected curves as seen in Figures 1 and 2.

\begin{table}
\label{simulations}
\begin{tabular}{|c|c|c|c|}
  \hline
  $k$ & $N$ & simulation result & Space-time estimate  \cr
  \hline \hline
0.5 & 100 & 18.1 & 22.8 \cr \hline
0.5 & 200 & 26.8 & 32.2 \cr \hline 
2.0 & 100 & 23.0 & 32.5 \cr \hline
2.0 & 200 & 34.7 & 45.8 \cr \hline
5.0 & 100 & 29.0 & 47.5 \cr \hline
5.0 & 200 & 44.9 & 66.9 \cr \hline
\end{tabular}
\caption{A comparison of space-time model estimates with the average
over 1000 simulations of boarding time results}
\end{table}
We also note that in all cases the Lorentzian estimate is larger.
 
Let $L_N$ be the random variable representing the boarding time
according to the boarding process model (the simulation results). We
define the discrepancy random variable $\Delta_N=L_N-\sqrt{N}d(k)$,
which measures the difference between the boarding time and the
Lorentzian estimate. The curvature of the metric given in
(\ref{nopolicy}) vanishes, and therefore we can apply a coordinate
transformation $W$ which changes the metric to the form
\begin{equation} \label{standard}
d\tau^2=4dxdy.
\end{equation}
As noted previously, in this case past-future causal chains correspond to
increasing sequences of points. We also note that, since the spacetime
points are sampled according to the volume form, they are uniformly
distributed. The discrepancy $\Delta_N$ has been studied in the
context of increasing subsequences of uniformly distributed points in a
rectangle with sides parallel to the coordinate axis \cite{BDJ}, and for a
right angle triangle with sides parallel to the coordinate axis \cite{BR}.
In both cases,
the discrepancy has order of magnitude $N^{1/6}$. In the case of the
rectangle the normalized discrepancy $\frac{\Delta_N}{N^{1/6}}$ is
given asymptotically by the Tracy-Widom distribution $F_2$ \cite{TW},
which measures the normalized discrepancy of the largest eigenvalue of
an $N\times N$ matrix in the Gaussian unitary ensemble (GUE) \cite{Mh},
in comparison with $2\sqrt{N}$. For $N/2$ uniformly distributed points
in a right angle triangle with sides parallel to the axis, the
normalized discrepancy is given asymptotically by the Tracy-Widom
Distribution $F_4$ \cite{TW}, which is the normalized discrepancy of
the largest eigenvalue in the Gaussian symplectic ensemble (GSE).
The averages for these distributions are $E(F_2)=-1.77$ and $E(F_4)=-2.3$.

Let $A=W(0,0)$ and $B=W(1,1)$. Let $U$ be the image under $W$ of the
unit square. Let $R$ be the rectangle with sides parallel to the $x$
and $y$ axis and corners $A$ and $B$, and denote by $T$ the above
diagonal triangle in $R$. $U$ is contained in $R$, and when $k<\ln 2$
it contains the above diagonal triangle $T$. The ratio of volumes
$Vol(R)/Vol(U)$ equals $\frac{e^k-1}{k}$ for all $k$, and thus, applying the
estimates from \cite{BDJ} and \cite{BR}, we see that for $k\leq \ln 2$,
as $N$ becomes large we have
\begin{equation}
\label{refined1}
F_4(z)\leq P(\Delta_N \leq (\frac{e^k-1}{k}N)^{1/6}z)() \leq F_2(z)
\end{equation} 
In particular, we obtain the finer estimate
\begin{equation}
\label{refined2}
-2.3\left(\frac{e^k-1}{k}N\right)^{1/6}
    \leq  E(\Delta_N)\leq -1.77\left(\frac{e^k-1}{k}N\right)^{1/6}
\end{equation}
Looking at the results from Table II with $k=0.5$, we see that the
refined estimate (\ref{refined2}) holds already for the realistic values
$N=100,200$; indeed $17.7<18.1<18.8$ for $N=100$ and $26.4<26.8<27.8$
for $N=200$. When $k>\ln 2$ the maximal curve contains a portion of
the bottom edge of the unit square. Using the methods of \cite{Jo}, it
can be shown that in such cases $\vert \Delta_N\vert >>N^{1/6}$.
However, we do not know how to compute analytically the order of
magnitude of the error. Figure 3 shows the behavior of $\Delta_N$ for
$k=3$ in Log-Log coordinates. As can be seen, the graph is essentially
linear, and linear regression suggests the formula
%\begin{equation}
%E(\Delta_N)=-6.78N^{0.227}
%\end{equation}
%\begin{figure}
\begin{equation}
E(\Delta_N)\sim -4.85N^{0.222}\,.
\end{equation}
\begin{figure}
\includegraphics[width=2.25in,height=2.25in]{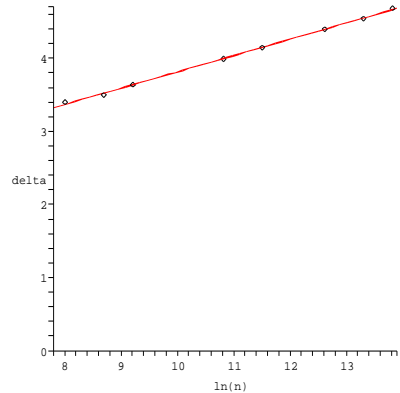}% Here is how to import EPS art
\caption{\label{fig:epsart3} Average discrepancy for $k=3$ 
as a function of $N$ in Log-Log scaling.}
\end{figure}

The spacetime metrics given by (\ref{model}) can be used for
comparing different boarding policies. Boarding policies such as
``passengers from row 40 and above board first, followed by the rest of
the passengers'' effect the passenger distribution function $p(q,r)$.
We compared the results of the spacetime computations (without finer
asymptotic corrections) with the results of detailed event driven
simulations of boarding processes \cite{VB,VVH}, which compare
different boarding policies. We computed spacetime based estimated
boarding times for 25 different boarding policies with parameter $k=4$.
Results based on detailed trace driven simulations for the
same policies are reported in \cite{VB}. A comparison of the results
shows that the spacetime estimates are in almost complete agreement
with the trace driven simulation results regarding the ranking of the
different policies. The correlation factor between the 25-dimensional
boarding time vectors is $0.97$. This is somewhat surprising given that
the trace driven simulations in \cite{VB} take into account many
details of actual boarding processes which are not considered by our
boarding process simulation. These include, among others, walking speed
of passengers, passengers sitting at the wrong row and full overhead
bins. The large discrepancies noted previously between
the boarding simulation and the spacetime estimates are less of a
factor, since when comparing boarding strategies only ratios of boarding
times matter and these are less affected by the discrepancies.
%A detailed account of the relevant computations and comparisons is given
%in \cite{BBSS}.
The main findings regarding actual boarding policies are:

- The commonly practiced back-to-front boarding policies, which attempt
to board passengers from the back of the airplane first, are ineffective
for realistic values of $3<k<5$. The intuition behind this statement
can be seen by the following simple reasoning. Assume the airline is
perfectly successful in enforcing back-to-front boarding and thus
passengers from the last row $m$ board first, followed by the
passengers from row $m-1$ and so on. If $k<1$, then passengers from all
rows can sit concurrently without interference since they do not block
each other. If $k>1$ then the passengers in row $m-1$ have to wait
until at least some of the passengers from row $m$ have sat down, and
similarly passengers from row $m-j-1$ have to wait for passengers from
row $m-j$ to sit. This leads to a linear sized chain of blocking, which is much larger than the $\sqrt{N}$-sized chains in random
boarding.

- Among row dependent policies which do not severely constrain
passengers, random boarding (no policy) is almost optimal.

- One can improve any row dependent boarding policy (including random)
by first allowing window seat passengers to board, followed by middle
seat and finally aisle passengers. Such policies lower the delay
parameter $D$, which affects the metric via scaling.

There are other discrete random processes which can be modeled via
2-dimensional Lorentzian geometry in a similar manner. One example of
interest is the Polynuclear growth (PNG) model, which is a particularly
simple $(1+1)$-dimensional growth process \cite{PS,Me}. The construction of
the present paper can be considered as generalizing the mapping of
Prahofer and Spohn \cite{PS} of the basic PNG model to permutations.
In this mapping the height of the PNG droplet corresponds to the length
of the longest increasing subsequence in a random permutation. In terms
of the airplane boarding process this corresponds to setting $k=0$.
More generally the methods of this paper provide a description of the
macroscopic shape of a PNG droplet in an environment with non-uniform
nucleation rates and lateral speeds.

A different example is the process of scheduling I/O requests to a disk
drive \cite{Ba,ABZ}.

In conclusion, we have shown that the airplane boarding process can be
asymptotically modeled by a boarding parameter dependent two
dimensional spacetime. The model can be used to analyze the dependence
of boarding times on the various boarding parameters and boarding
policies. The discrepancy between the asymptotic model and finite
population results is closely related to RMT, at least for thin
passengers. The analysis carries applications to the design of good
airplane boarding policies.

We are grateful to Perci Deift, Ofer Zeitouni and Jinho Baik for very
useful discussions.
%We may also consider the fluctuations in boarding time around the first order approximation given by 
%$d(M)\sqrt{n}$. We may phrase an essentially
%equivalent problem as follows:
%
%Given a bounded 2 dimensional causal space-time, what are the fluctuations in the size of the longest chain with respect to the causal
%structure among $n$ points sampled according to the volume form.

%%%%%%%%%%%%%%%%%%%%%%%%%%%%%%%%%%%%%%%%%%%%%%%%%%%%%%%%%%%%%%%%

\begin{thebibliography}{99}

\bibitem{MMM} S. Marelli, G. Mattocks and R. Merry, Boeing Aero Magazine {\bf 1}, 
PAGE NUMBER (2000).
%\bibitem{So} L. Bombelli, J. Lee, D. Meyer and R.D. Sorkin, Phys Rev. Lett. {\bf 59}, 521 (1987).

\bibitem{VB} H. Van Landegham and A. Beuselinck, Euro. J. of Op. Res. {\bf 142}, 294 (2002).

\bibitem{VVH} M. van den Briel, J. Villalobos and G. Hogg, Proc. of IERC (CD ROM only, 2003).

\bibitem{Ma}
C.L. Mallows, Bull. Inst. Math. Appl. {\bf 9}, 216 (1973).

\bibitem{AD} D. Aldous and P. Diaconis,
Bull. Amer. Math. Soc. {\bf 36}, 413 (1999).

%\bibitem{Fr}
%M.L. Fredman, Disc. Math., {\bf 11}, 29 (1975).

\bibitem{DZ}
J.D. Deuschel and O. Zeitouni, Ann. of Prob. {\bf 23}, 852 (1987).

\bibitem{VK} 
A. Vershik and S. Kerov, Soviet Math. Dokl. {\bf 18}, 527 (1977).

%\bibitem{Ha}
%J.M. Hammersley, Proc. Sixth Berkeley Symp. Math. Statist. and prob. {\bf 1}, 345 (1972).
\bibitem{BDJ} J. Baik, P. Deift and K. Johansson, J. Amer. Math. Soc. {\bf 12}, 1119 (1999).

\bibitem{BR} J. Baik and E. Rains, Duke J. of Math {\bf 109}, 205 (2001).

\bibitem{TW} C.A. Tracy and H. Widom, Commun. Math. Phys. {\bf 159}, 151 (1994); {\bf 177}, 727 (1996). 

\bibitem{Mh} M.L. Mehta, {\it Random Matrices}, (Academic press, 2004).

\bibitem{Jo} K. Johansson, Commun. Math. Phys. {\bf 209}, 437 (2000).

\bibitem{PS}
M. Prahofer and H. Spohn, Phys. Rev. Lett. {\bf 84}, 4882 (2000).

\bibitem{Me}
P. Meakin, {\it Fractals, Scaling and Growth Far from Equilibrium} (Cambridge University Press, Cambridge, England, 1998).

%\bibitem{KPZ} M. Kardar, G. Parisi and Y.-C. Zhang, Phys. Rev. Lett. {\bf 56}, 889 (1986).

\bibitem{ABZ}
M. Andrews, M. Bender and L. Zhang, Algorithmica {\bf 32}, 277 (2002).

\bibitem{Ba} E. Bachmat, Proc. of Symp. Th. of Comp., 277 (2002).



\end{thebibliography}
\end{document}